# Amy Ogle
## Isobel Falconer

"My best man is Ogle," declared an examiner for Cambridge's Natural Sciences Tripos, in 1876. He was unaware that "Ogle" was a woman.[1]

Amy Ogle (1848-1878) came top in the Natural Sciences Tripos (NST) 14 years before Philippa Fawcett achieved a similar distinction in Cambridge's Mathematical Tripos. Fawcett is famous, while Ogle is unknown, reflecting the much greater prestige of mathematics over natural sciences at the time, and how the position of women had changed in those 14 years.

When Amy took the NST, women required special permission to take Tripos exams and their results were not published as the men's were. Evidence of her achievement has remained hidden in the NST mark books, apart from a brief mention in an obscure government report two years later. Her college, Newnham, *did* know that she had gained a First, from the customary informal report sent to them by the examiners; this was leaked to the press causing a brief stir. The *Athenaeum* reported,

> During the recent Natural Science Tripos examination at Cambridge, a lady, Miss Ogle, who is a student at Newnham Hall, the Cambridge college for women, was, by the permission of the examiners, subjected to precisely the same examination, as that which the members of the University underwent. She acquitted herself in such a manner, as would have entitled her, had she been an undergraduate, to a place in the first-class.[2]

---

[1] William Grylls Adams to Millicent Fawcett, 7 February 1896. Manchester Central Library, Letters to Mrs Fawcett, GB127.M50/3/1/20
[2] *The Athenaeum*. (6 January 1877), p.19.

Figure 1: NST mark book for 1876. As was usual, women's marks were added as an afterthought at the bottom of the page (Courtesy Cambridge University Library)

Amy took six sciences in Part II, not an unusual number, coming top in zoology, second in physiology and third in botany. However, what intrigued me more was that she was the only woman in the Maxwellian era of the Cavendish to obtain a respectable mark in physics. Her modest 50 paled in comparison with the 686 of the top candidate – William Napier Shaw, the meteorologist. But 50 was still high enough to place Amy 6th out of the 16 people (15 men) taking Part II physics.

In 1874 the Cavendish had opened its doors to "any member of the University". But women were not members, and not until 1878, two years after Amy took the Tripos, did Maxwell reluctantly permit the Demonstrator, William Garnett, to admit women for an accelerated course during the Long Vacation. So, when, and how did Amy learn her physics?

Miss Clough, Principal of Newnham, tells us that Amy "received her early education under her father's special superintendence."[3] This was probably along with the boys at the small boarding school that her parents, John and Sarah Ogle, ran in a rented country house, "St Clere", in Kent. They were strongly influenced by Pestalozzi and a belief that "the family must be the model of the [educational] community we form. It is a Divine Institution."[4] A corollary was mixing between the schoolmaster's family and pupils, and John's unconventional stand that teaching boys and girls together was the best way to ensure their social and moral development. The Pestalozzian emphasis on observation would have prepared Amy well for success in science and she may have gained a good mathematical

---

[3] Anne J. Clough, Correspondence, *Journal of the Women's Education Union*, Feb 15, 1878, vol 6, issue 62, pp30-31

[4] John Ogle (1871)" The application of theory in the practice of education" Lecture to the College of Preceptors, 14 June 1871, *Educational Times* 24(123), 71-78 (71)

grounding from the succession of teachers at St Clere that included a Cambridge wrangler (i.e. a first class in the Mathematics Tripos) Robert Tucker. Certainly at least three pupils from the school, including Amy's elder brother John Lockhart, became Cambridge wranglers.[5]

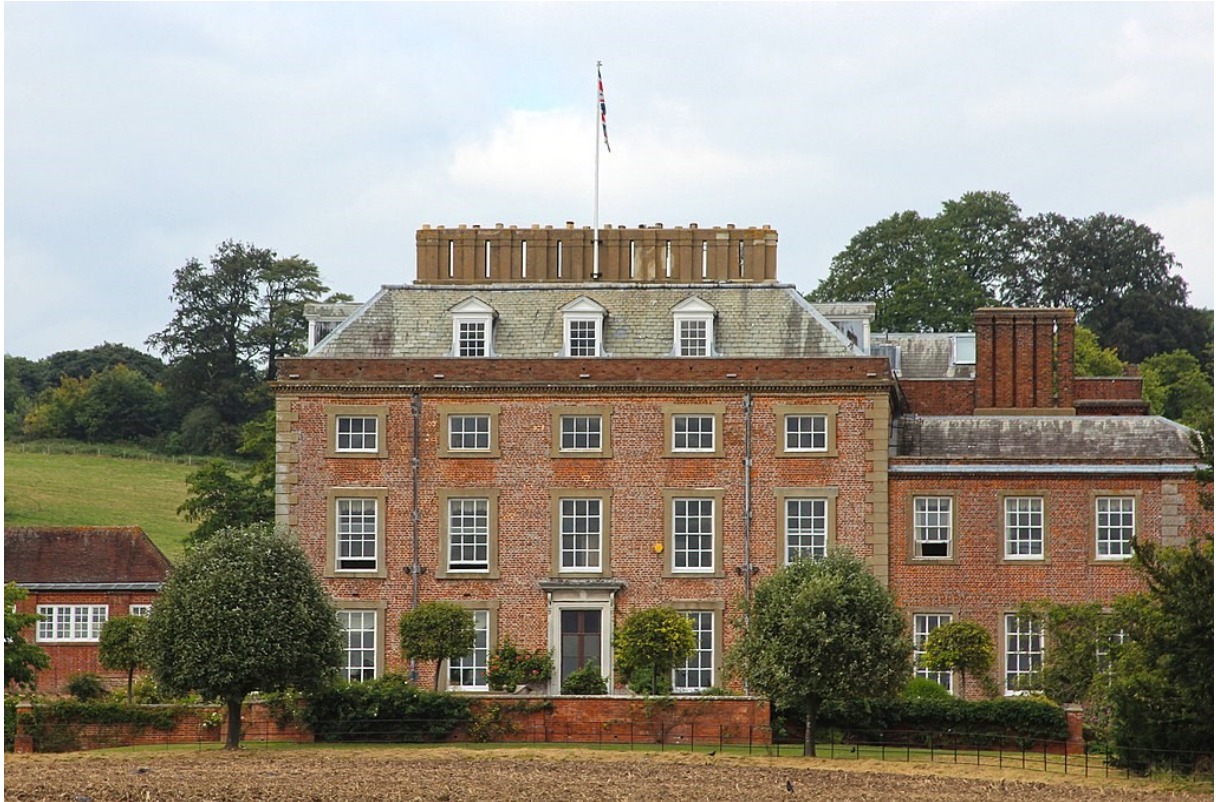

Figure 2: St Clere, which Amy's parents rented to house their boarding school (Oast House Archive / *St Clere* / CC BY-SA 2.0)

Miss Clough characterised Amy as "industrious" and "earnest", though "interesting",[6] but did not remark on how determined and quietly unconventional she must have been. By the 1871 census, when her sisters were all living with their parents, Amy was teaching at a girls' school in Edinburgh. Here she may have engaged with the nascent opportunities for women's higher education offered by the Edinburgh Ladies Educational Association, or the Watt Institute, although so far I have found no evidence.

Amy came to Newnham in 1873, two years after its foundation, when she was 25. She supported herself by scholarships, was one of two students living "out", perhaps because it was cheaper, and probably paid the reduced fees for women intending to become teachers. Although Maxwell barred them from the Cavendish, Philip Main, Superintendent of St John's College Laboratory, welcomed women and it seems likely that Coutts Trotter, Chairman of

---

[5] John Stuart Jackson, 5th wrangler 1851, William Previté, 32nd wrangler 1860; John Lockhart Ogle, 29th wrangler 1866; ACAD https://venn.lib.cam.ac.uk/; M J M Hill, Obituary Notice : Robert Tucker, *Proc. London Math. Soc.* (1905), xii-xx

[6] Clough, Blanche Athena. *A Memoir of Anne Jemima Clough.* London, New York : E. Arnold, 1897, p201-202; Anne J. Clough, op. cit. 3

Newnham's Council, admitted them to his physics lectures at Trinity. But Amy also had another opportunity – the short summer courses for teachers offered at South Kensington. These included "sound", "light", and "steam", which Amy may have attended, and biological subjects which she certainly did. In 1878 William Thiselton-Dyer reported to the Government Department of Science and Arts,

> "Miss A.H. Ogle attended my course in Botany in 1876 as a paying student. In the Michaelmas term of the same year she answered the papers set in the Natural Sciences Tripos at Cambridge, where I was one of the examiners. Although the regulations do not allow the results in the case of women to be officially published, Miss Ogle's marks would have placed her at the head of the tripos, and she did particularly well in Botany. Miss Ogle is now, I believe, head of a large middle-class school."[7]

Unfortunately, Thiselton-Dyer was mistaken in his belief; Amy had died in childbirth a few months earlier. Following her Tripos success she lectured in Natural Sciences at Newnham for a term before marrying an evangelical Christian Jew, Dr Joseph Koppel. But marriage and incipient motherhood did not curtail her career aspirations. In November 1877, already pregnant, she was appointed Principal of the Training College being established in London's Bishopsgate by the Teachers' Training and Registration Society.[8] Her daughter Una was born on 13 January 1878 and Amy died the same day, before taking up her appointment.

Amy was among the first of a group who, over the next 40 years, forged career paths into academia for women via prestigious appointments at teacher training colleges.[9] But, in negotiating to continue her career while married with young children, she started on a different path to that that subsequently became stereotypical, of spinster academics.[10] Had she lived, might her example have prompted alternative models?

---

[7] W. Thiselton-Dyer, quoted by Lieut-Col. Donnelly, 'Twenty fifth report of the Science and Art Department of the Committee of Council on Education, with appendices: Appendix B: Science Instruction', Parliamentary Papers, Command Papers, 1878, p34

[8] Subsequently Maria Gray College, which merged with the West London Institute of Higher Education in 1976, and was absorbed into Brunel University in 1995.

[9] Dyhouse, Carol. *No Distinction Of Sex?: Women In British Universities, 1870-1939*. Routledge, 2016, p136

[10] Dyhouse, p161, notes that 79-85% of female academics in the 1884-1904 period remained lifelong spinsters